\PassOptionsToPackage{unicode}{hyperref}
\PassOptionsToPackage{hyphens}{url}
\documentclass[
  10pt,
]{article}
\usepackage{xcolor}
\usepackage[margin=1in]{geometry}
\usepackage{amsmath,amssymb}
\usepackage{iftex}
\ifPDFTeX
  \usepackage[T1]{fontenc}
  \usepackage[utf8]{inputenc}
  \usepackage{textcomp} 
\else 
  \usepackage{unicode-math} 
  \defaultfontfeatures{Scale=MatchLowercase}
  \defaultfontfeatures[\rmfamily]{Ligatures=TeX,Scale=1}
\fi
\usepackage{lmodern}
\ifPDFTeX\else
\fi
\IfFileExists{upquote.sty}{\usepackage{upquote}}{}
\IfFileExists{microtype.sty}{
  \usepackage[]{microtype}
  \UseMicrotypeSet[protrusion]{basicmath} 
}{}
\makeatletter
\@ifundefined{KOMAClassName}{
  \IfFileExists{parskip.sty}{%
    \usepackage{parskip}
  }{
    \setlength{\parindent}{0pt}
    \setlength{\parskip}{6pt plus 2pt minus 1pt}}
}{
  \KOMAoptions{parskip=half}}
\makeatother
\usepackage{longtable,booktabs,array}
\usepackage{calc} 
\usepackage{etoolbox}
\makeatletter
\patchcmd\longtable{\par}{\if@noskipsec\mbox{}\fi\par}{}{}
\makeatother
\IfFileExists{footnotehyper.sty}{\usepackage{footnotehyper}}{\usepackage{footnote}}
\makesavenoteenv{longtable}
\usepackage{graphicx}
\makeatletter
\newsavebox\pandoc@box
\newcommand*\pandocbounded[1]{
  \sbox\pandoc@box{#1}%
  \Gscale@div\@tempa{\textheight}{\dimexpr\ht\pandoc@box+\dp\pandoc@box\relax}%
  \Gscale@div\@tempb{\linewidth}{\wd\pandoc@box}%
  \ifdim\@tempb\p@<\@tempa\p@\let\@tempa\@tempb\fi
  \ifdim\@tempa\p@<\p@\scalebox{\@tempa}{\usebox\pandoc@box}%
  \else\usebox{\pandoc@box}%
  \fi%
}
\def\fps@figure{htbp}
\makeatother
\NewDocumentCommand\citeproctext{}{}

\makeatletter
 \let\@cite@ofmt\@firstofone
 \def\@biblabel#1{}
 \def\@cite#1#2{{#1\if@tempswa , #2\fi}}
\makeatother
\newlength{\cslhangindent}
\newlength{\csllabelwidth}
\newenvironment{CSLReferences}[2] 
 {\begin{list}{}{%
  \setlength{\itemindent}{0pt}
  \setlength{\leftmargin}{0pt}
  \setlength{\parsep}{0pt}
  \ifodd #1
   \setlength{\leftmargin}{\cslhangindent}
   \setlength{\itemindent}{-1\cslhangindent}
  \fi
  \setlength{\itemsep}{#2\baselineskip}}}
 {\end{list}}
\usepackage{calc}

\providecommand{\tightlist}{%
  \setlength{\itemsep}{0pt}\setlength{\parskip}{0pt}}
\usepackage{bookmark}
\IfFileExists{xurl.sty}{\usepackage{xurl}}{} 
\makeatletter
\@ifundefined{xmpquote}{}{}
\makeatother
\hypersetup{
  pdftitle={How Far Can Chord-Symbol Time-Series Adaptation Carry Genre Identity?},
  pdfauthor={Jinju Lee --- PearlLeeStudio},
  hidelinks,
  pdfcreator={LaTeX via pandoc}}

\title{How Far Can Chord-Symbol Time-Series Adaptation Carry Genre
Identity?}
\usepackage{etoolbox}
\makeatletter
\providecommand{\subtitle}[1]{
  \apptocmd{\@title}{\par {\large #1 \par}}{}{}
}
\makeatother
\subtitle{Capabilities and Boundaries in Multi-Genre Chord-Symbol
Modeling}
\author{Jinju Lee --- PearlLeeStudio}
\date{June 2026}

\begin{document}
\maketitle

\subsection{Abstract}\label{abstract}

Harmony is a compact symbolic layer where mathematical pitch relations,
acoustic consonance, and musical convention meet. This report studies
that layer directly: chord-symbol sequences are treated not as a
complete representation of music, but as an interpretable and
controllable time series for genre-local harmonic modeling. The starting
point is a frozen Music Transformer base checkpoint, released as the
endpoint of a pop-jazz rehearsal-mix fine-tune but verified in this
revision to be weight-identical to the Phase-0 pop baseline, so
that all adaptation gains are measured over a pop-dominated harmonic prior
(see Changes in v2). From this base, I evaluate how far small
adaptation interfaces
can extend the model to eleven target genres: blues, bossa nova, Bach
chorales, country, electronic, folk, funk, gospel, hip-hop, R\&B/soul,
and rock. The main evaluation compares
five adaptation methods (LoRA, IA3, BitFit, prefix tuning, and full
fine-tuning) over 11 genres and 3 seeds, yielding a complete 165-cell
grid. All five methods improve over the frozen base on held-out chord
prediction, with macro gains from +2.89 to +3.61 percentage points.
Top-1 also scores the three header tokens that open every sequence (key,
time signature, genre marker), whose prediction moves a whole run at
once: only a matched adapter predicts its genre marker, and whether a
run predicts the placeholder C-major key that every Chordonomicon song
carries is close to arbitrary. Scored without them, the macro gains
are +1.64 to +2.38 pp, and outside Bach chorales LoRA, IA3, prefix
tuning, and full fine-tuning give the same gain (+1.30 to +1.34 pp).
That gain reflects a handicap of the frozen base: for every target
genre but rock it reads a genre marker that it never read in training,
and given the pop marker instead it comes within 0.12 pp of the released
matched LoRA adapter in every non-chorale genre. Only on Bach chorales
does adaptation add clearly more (+9.6 pp for LoRA over the base given
the pop marker). LoRA
and IA3 have the highest macro top-1 scores, but pairwise Wilcoxon tests
with Holm and Benjamini-Hochberg correction do not support a decisive
method winner. A matched-data-size control (ten genres, four methods) agrees: at a
common corpus size the methods differ only by whole key-token steps, and
without the header tokens their gains agree within 0.03 pp. A control-token baseline is also strong, and wrong-genre adapters
often beat the frozen base as scored, though only where it reads an
untrained marker, suggesting that much of the adaptation effect is
exposed through lightweight conditioning over a reusable harmonic base
rather than through one particular adapter family or purely
genre-specific adapter memory. Outside Bach chorales, the matched
adapter's edge over the best wrong-genre adapter is within 0.2
percentage points once the header tokens are not scored. Additional diagnostics, including rank
sweeps, wrong-genre adapter rotation, a matched-data-size control, a
base-checkpoint ablation (in v2, a same-weights control), chord-only
genre classification,
generated-output distribution statistics, real-song chord-chart
evaluation, and duplicate/near-duplicate analysis, support a bounded conclusion: chord-symbol adaptation clearly improves
genre-local harmonic prediction for Bach chorales, but for the ten
Chordonomicon-derived genres it adds little beyond a usable genre
marker, and chord symbols alone do not carry complete genre
identity. The report therefore avoids claims about perceived genre
authenticity or full musical quality. Those require controlled listener
or musician evaluation.

\begin{center}\rule{0.5\linewidth}{0.5pt}\end{center}

\subsection{Changes in v2}\label{changes-in-v2}

This revision corrects the description of the frozen base checkpoint.
No reported number changes. What changes is what the released base
artifact is and how the base-checkpoint ablation must be read.

\begin{itemize}
\tightlist
\item
  \textbf{Verified weight identity.} A full SHA-256 comparison over all
  25,841,152 parameters\footnote{The SHA-256 covers the 25,841,152-element serialized state dict. The model has 25,661,440 unique parameters, the difference being the tied input/output embedding ($351\times512$) stored as two tensors.}, run on independent hardware (2026-06-11),
  shows that the released F1 base checkpoint
  (\texttt{ft\_jazz\_pop80/best.pt}, Hugging Face
  \texttt{TheArtist-MusicTransformer-ft-pop80}) is weight-identical to
  the Phase-0 pop baseline checkpoint.
\item
  \textbf{Mechanism.} Two factors combined. The F1 run resumed from the
  Phase-0 checkpoint together with its optimizer and scheduler state,
  whose learning rate had decayed to 0, so its first fine-tune epoch
  (epoch 3) updated no weights and produced a candidate checkpoint
  identical to Phase 0. The configured fine-tune learning rate and
  warmup never applied. From epoch 4 the run trained at the restored
  base rate of $3 \times 10^{-4}$ under its cosine schedule, and its
  training log (\texttt{eval\_results.csv}) shows held-out jazz top-1
  rising from 72.86 at epoch 3 to a peak of 81.03 at epoch 6. But
  best-checkpoint selection minimized validation loss on a
  pop-dominated validation mix, which rose monotonically during jazz
  fine-tuning (0.5578 $\rightarrow$ 0.5901), so the unchanged epoch-3
  checkpoint, the Phase-0 weights, was retained as ``best''.
  The genuinely fine-tuned epoch checkpoints existed only on the
  training machine and were not released.
\item
  \textbf{Consequences.} (a) All adaptation gains over the frozen base
  are gains over a \textbf{pop-dominated harmonic prior}. The substantive
  findings are unaffected and the interpretation is cleaner. All
  released evaluation numbers were produced with the served artifacts
  and reproduce on independent CPU hardware to within 0.02 pp at the
  model level. (b) The base-checkpoint ablation (Section 6.9) is
  retroactively a \textbf{same-weights control}, and its ``essentially
  unchanged'' result independently corroborates the weight identity
  (compared on one split, the two runs differ by at most 0.13 pp in macro
  top-1, so the $\pm$0.4 pp quoted here before v5 was not a run-to-run
  noise estimate). (c) The v1 claim that the F1 base ``retains a richer
  harmonic vocabulary'' than the Phase-0 pop base is \textbf{withdrawn}:
  identical weights imply identical sampling distributions. The
  author's design intent, a jazz-leaning base chosen for harmonic
  character, stands as intent, but the released artifact did not embody
  it.
\item
  \textbf{Scope.} F2--F5 checkpoints are unaffected (weights all
  distinct, hash-verified). The 11 per-genre adapters were trained on,
  and are self-consistent with, the released base.
\item
  \textbf{Meta-observation.} This is itself an instance of the
  phenomenon the project studies: metric-optimal checkpoint selection
  on a mismatched validation distribution can silently discard
  adaptation.
\item
  \textbf{Released-adapter rank selection (disclosure).} The per-genre
  LoRA ranks used for this report's selected-rank evaluation were
  chosen by highest validation top-1, while the released adapters on
  Hugging Face were deliberately selected by lowest validation loss
  with a top-1 tiebreak, consistent with training-time best-checkpoint
  selection. The two criteria pick different ranks for six genres.
  Section 5.4 now documents this so that reproduction from the released
  adapters targets the released-rank cells of the rank sweep rather
  than the selected-rank row.
\end{itemize}

\subsection{Changes in v3}\label{changes-in-v3}

No reported conclusion changes. (i)~A suitable jazz-adapted base now
exists: the selection-corrected retrain of the F1 slot (jazz-only
validation selection, released as \texttt{ft-pop80-v2}) is hash-distinct
from the Phase-0 baseline (jazz top-1 75.05 on the unified 9-source
test). The selection rule reproduces across three random seeds in
matched-data retrains on that corpus (jazz top-1 $75.76 \pm 0.03$), so the
base-robustness ablation (Sections 6.9 and 8) is now gated by
experimental effort rather than checkpoint availability. (ii)~Several
reported summary statistics were corrected for exact data-faithfulness
against the released CSVs: the Section 6.3 macro, median, and non-chorale
LoRA gains and the minimum per-genre mean gain (all taken from the
selected-rank three-seed grid summary \texttt{lora\_multi\_seed.md}), and
the Section 6.9 per-genre count and macro magnitudes. No ranking,
significance result, or qualitative conclusion changes. (iii)~Section 5.4 now states that the
selection-corrected jazz retrains and their multi-seed reproduction were
trained on cloud Tesla~T4 GPUs (Colab/Kaggle) rather than the laptop GPU
used for the main grid, all in fp16 mixed precision and unified on a
common full-precision CPU re-evaluation. A parameter-count footnote
clarifies the serialized-element count versus the model parameter total.

\begin{center}\rule{0.5\linewidth}{0.5pt}\end{center}

\subsection{Changes in v4}\label{changes-in-v4}

One description in Section~5.2 is corrected. The trainable-footprint
table previously listed LoRA at the $r = 32$ configuration (1,154,048
parameters, 4.5\%) as if the method had a single footprint. The main
grid selects the rank per genre (4--64), so the LoRA footprint ranges
from 465,920 ($r = 4$) to 1,940,480 ($r = 64$) parameters, 1.8--7.6\%
of the model, and the adjacent ``0.9--4.5\%'' range becomes
``0.9--7.6\%''. Additionally, the v3 note and Sections~6.9 and~8 now
attribute the three-seed $75.76 \pm 0.03$ reproduction to matched-data
retrains on the 9-source corpus: the released \texttt{ft-pop80-v2}
itself (trained on the 6-source corpus) scores 75.05 on that test. v4
also corrected four data-profile and table values that this note did
not list (see Changes in v5). No conclusion changes.

\begin{center}\rule{0.5\linewidth}{0.5pt}\end{center}

\subsection{Changes in v5}\label{changes-in-v5}

v5 adds two measurements that change the reading of the results, and
corrects descriptions. No logged value or released artifact changes.
(i)~Top-1 scores every target token, including the
three header tokens that open each sequence (key, time signature, genre
marker), and a model makes nearly the same header prediction for every
test sequence, so each header target moves a whole run at once (Section
5.3). F1 never predicts a target genre's marker and a matched adapter
always does. Chordonomicon has no key annotation, so every song of the
ten non-chorale test sets carries the placeholder
\texttt{{[}KEY:Cmaj{]}}, while 12-key augmentation leaves a model's key
prediction after \texttt{{[}BOS{]}} close to uniform: four of the eleven
released adapters (blues, bossa nova, country, gospel) predict C major
and F1 does not, and the runs without kept checkpoints fall on score
levels one key step apart, apart from nine BitFit cells (Section 5.3).
The abstract and Sections 5.3, 6.1, 6.3, 6.4, 6.10, 7.3, and 9 now also
report results without the header targets. The macro gains become +1.64
to +2.38 pp, and outside Bach chorales LoRA, IA3, prefix tuning, and full
fine-tuning give the same gain (+1.30 to +1.34 pp). The method
differences of the matched-data-size control and the rank preferences of
the rank sweep are whole key steps, so those readings are withdrawn.
Outside Bach chorales the matched adapter's margin over the best
wrong-genre adapter falls from 0.82 to 1.25 pp to at most 0.20 pp.
(ii)~F1 reads every target genre's marker except rock's as an input that
it never read in training (Section 5.3). Rescored on CPU with the pop
marker in that position, F1 comes within 0.12 pp of the released matched
LoRA adapter without the header targets in every non-chorale genre,
and in the five genres checked also on test songs that are not in its
training split, so outside Bach
chorales the gain of (i) reflects that handicap rather than added
harmonic knowledge. The abstract and Sections 1, 5.3, 6.1, 6.2, 6.4,
6.5, 6.7, 6.10, 6.12, 7, and 9 now say so. The wrong-genre adapters
beat F1 only where it reads an untrained marker, and Section 6.1 now
traces rock's near-zero gain to its familiar marker rather than to the
overlap of (iii), which every genre shares. (iii)~Sections 4.1 and 4.3
now state that 79.7 to 82.8\% of each non-chorale target test set also
appears, with the same chord sequence, in the Phase-0 training split of
the frozen base, because the target subsets were drawn later from the
same Chordonomicon file. Section 4.3 also gives the share of test songs
that repeat a song of their own subset's training split (2.16\% on
average, 18.4\% for Bach chorales), where earlier versions gave the
share of training sequences that recur in test (0.28\%) as the
train-to-test duplicate rate. (iv)~The base-checkpoint control of Section
6.9 had compared a Phase-0 validation-split macro with an F1 test-split
macro. On one split the two runs differ by at most 0.13 pp, so the
earlier $\pm$0.4 pp was not a noise estimate, and the three-seed spread
of the method macros (0.09 to 0.32 pp) takes its place in Sections 5.1,
6.9, and 7.3. (v)~The F1 weight identity had a second cause that
earlier versions did not state: the F1 run resumed with the Phase-0
optimizer and scheduler state, whose learning rate had decayed to 0, so
its first fine-tune epoch changed no weights and produced a candidate
identical to Phase 0, which selection on the pop-dominated mixed
validation loss then kept. The configured fine-tune learning rate and
warmup never applied (Changes in v2 and Section 4.1). (vi)~Method
descriptions now match the code. Every method, the control-token
baseline included, trains the full token embedding table and output
projection, and the control-token baseline trains nothing else. The
BitFit and full fine-tuning footprints are those of training (413,184
and 25,849,344 parameters), and the percentages are now taken of the
25,849,344-parameter adapted model, so LoRA spans 1.8 to 7.5\%. The
Phase-0 base is pop-dominated rather than pop-only, its markers come
from Chordonomicon's genre tags mapped onto pop, rock, and jazz, eight
of the eleven targets add a new genre marker, all ten non-chorale genres
come from Chordonomicon tags and the bossa nova slice pools bossa,
samba, latin, salsa, and cumbia tags (Section 4.1), the
matched-data-size control covers ten genres and four methods, and
Section 5.4 notes how the full fine-tuning and BitFit cells were
reloaded for testing, which most likely explains BitFit's low cells in
the three largest corpora. (vii)~Section 6.2 marks the control-token
macro as a validation-split value and adds F1's validation-split macro,
and Section 5.4 limits the CPU re-evaluation
to the Phase-0 baseline and the released retrain (the v3 note had
described a common CPU re-evaluation of all retrains, whose three-seed
numbers are training-log values). Sections 4.4, 5.4, and 8 describe the
release as it stands (weights on Hugging Face, code on GitHub, split
files not distributed), Section 4.4 gives the SHA-256 of the source
file, Section 6.3 notes where its three-seed summary differs from the
main grid, Section 6.4 notes which rotation rows the released adapters
reproduce, Section 6.9 notes which logs its comparison reads, Section
6.10 notes the epoch budget of its runs, Figure 4 is redrawn with rock
at the training size of its subset, Figures 2 and 3 now appear in the
order of the numbers printed on them, and one reference is corrected.
(viii)~v4 also made four corrections that its
note did not list: the Table 1 rock row (the 39,128 training sequences
of the rock subset instead of all 152,509 training sequences that carry
the rock marker, with the row's other statistics recomputed), the
corpus-size range in Section 4.1 (296 Bach-chorale to 49,388 country
sequences), the Section 4.3 duplicate rates (means 0.28\% exact and
0.46\% key-normalized), and the bossa value in the Section 6.1
per-genre table (+3.55 pp, the three-seed mean).

\begin{center}\rule{0.5\linewidth}{0.5pt}\end{center}

\subsection{1. Introduction}\label{introduction}

This project began as an engineering problem. I was building an
interactive chord-composition system in which a user could edit a
harmonic skeleton, ask a model for continuations, and keep the result
under direct musical control. The system needed a practical capability:
a frozen chord model should be extensible to new genres without training
and serving a full separate model for every genre. That engineering need
became the empirical question of this report:

\begin{quote}
How much genre identity can chord-symbol time series carry, and where
does that representation reach its boundary?
\end{quote}

The question is narrower than full music generation. A genre is not only
a sequence of chords. It also includes rhythm, timbre, instrumentation,
voicing, production, performance practice, lyrical convention, and
listening context. Still, harmony is a meaningful intermediate layer. In
Western and jazz-adjacent traditions, chord symbols encode root motion,
functional patterns, modal mixture, cadential habits, extensions, and
repetition. These structures connect mathematical pitch relations,
acoustic consonance and dissonance, and the practical vocabulary
musicians use to compose and communicate.

This is why the work focuses on chord-symbol generation rather than
melody, lyrics, raw audio, or full arrangement. The choice is not a
legal claim that chord symbols eliminate copyright or licensing
concerns. They do not. Instead, this work studies a compact symbolic
representation of harmonic structure, using research-permitting
chord-chart corpora, and avoids modeling melody, lyrics, recordings, or
performer-specific audio. The representation is deliberately limited.
That limitation is the point: if chord symbols are used as a
controllable layer for music AI, we need to know both what they can
support and what they cannot.

The prior arXiv report in this project studied a pop-to-jazz
rehearsal-mix problem: how much pop data should be retained while
fine-tuning a pop chord model toward jazz
(\href{https://arxiv.org/abs/2605.04998}{arXiv:2605.04998}). The current
report changes the question. It freezes a base checkpoint from that
study, released as the pop-jazz F1 endpoint but verified in v2 to
be weight-identical to the Phase-0 pop baseline (Section 4.1),
and asks whether multiple adaptation methods can specialize it to
eleven target genres. This is not a PEFT comparison that was later given a
musical interpretation. The PEFT comparison follows from a
composition-tool problem: how to make genre-specific chord continuation
controllable while keeping one reusable base model. LoRA was the first
practical choice because it supports modular genre adapters, but the
study is not a ``LoRA for music'' paper. LoRA, IA3, BitFit, prefix
tuning, full fine-tuning, and a control-token baseline are used as
probes of the genre information available in chord-symbol sequences.

The central result is bounded. Every main method improves over the
frozen base on the held-out target-genre chord prediction task as it is
scored here, but outside Bach chorales the frozen base does as well once
it is given a genre marker it was trained on and the header targets are
set aside (Section 5.3). No method is statistically dominant after
correction, and several diagnostics show that chord-symbol genre
information is real but incomplete. This makes the contribution a
representation-boundary study rather than a method leaderboard.

The report makes four contributions:

\begin{enumerate}
\def\labelenumi{\arabic{enumi}.}
\tightlist
\item
  A complete 5-method $\times$ 11-genre $\times$ 3-seed evaluation of chord-symbol
  genre adaptation over a frozen base checkpoint (verified
  weight-identical to the Phase-0 pop baseline, Section 4.1).
\item
  A control-token and wrong-genre diagnostic package that tests whether
  gains are method-specific, genre-specific, or generic adaptation
  effects, yielding two reusable observations: a control-token baseline that
  tunes only the input and output layers comes close to adapter
  performance, and outside Bach chorales the frozen base matches the
  matched adapters once it reads a genre marker it was trained on, which
  also explains why many wrong-genre adapters beat it, while the matched
  adapter's edge over the best wrong one is small.
\item
  A chord-only genre classifier, transition-level diagnostics,
  generated-output statistics, and real-song chord-chart evaluation that
  check whether held-out top-1 gains correspond to broader chord
  behavior, plus a matched-data-size control that tests whether the
  method ranking is representational or an artifact of corpus size, and
  a base-checkpoint ablation that, with the base verified
  weight-identical to Phase 0, serves as a same-weights noise control
  (Section 6.9).
\item
  A conservative framing of chord symbols as a useful but bounded layer
  for controllable music AI, with musician-centered evaluation left as
  the required next step before perceptual claims.
\end{enumerate}

\begin{center}\rule{0.5\linewidth}{0.5pt}\end{center}

\subsection{2. Background}\label{background}

\subsubsection{2.1 Chord Symbols as a Time
Series}\label{chord-symbols-as-a-time-series}

A chord progression can be represented as a sequence of discrete
symbolic events: key markers, time-signature markers, bar markers, genre
markers, and chord tokens such as \texttt{C:maj7}, \texttt{A:min}, or
\texttt{G:7}. This representation discards voicing, rhythmic placement
finer than the chord grid, timbre, articulation, melody, and lyrics. It
keeps a compact abstraction of harmonic motion.

This compactness is useful for modeling. Compared with polyphonic MIDI
or audio, chord-symbol sequences have shorter contexts, smaller
vocabularies, and more interpretable errors. A next-chord prediction
error can be inspected harmonically: the model chose a diatonic
substitute, missed a secondary dominant, flattened a cadence, or
over-predicted a common loop. At the same time, top-1 and top-5 accuracy
are not musical-quality metrics. They measure agreement with corpus
continuations, not whether a progression is artistically better.

\subsubsection{2.2 Genre Identity Is
Layered}\label{genre-identity-is-layered}

Genre identity is not located in one musical layer. Some genres are
harmonically distinctive, while others may be defined more strongly by
groove, sound design, production, instrumentation, vocal style, or
performance practice. This is especially important for hip-hop,
electronic, funk, and many pop-adjacent genres, where harmonic
vocabulary may overlap heavily with other corpora while rhythmic and
timbral cues carry much of the identity.

The chord-symbol layer is therefore expected to carry partial
information. If adaptation gains are large and genre-specific, the layer
is useful. If gains saturate quickly, if a control token performs
similarly to adapters, or if a chord-only classifier has low macro F1,
the representation boundary is visible.

\subsubsection{2.3 Adaptation Interfaces as
Probes}\label{adaptation-interfaces-as-probes}

Parameter-efficient fine-tuning methods are usually discussed as
practical ways to adapt large models with fewer trainable parameters.
This report uses them differently. The methods are probes: if several
small interfaces can improve target-genre prediction over a frozen base,
the chord-symbol layer contains reusable genre-local information. If the
interfaces tie, trade wins, or show shallow capacity scaling, the result
says more about the representation and data than about a single best
method.

The adaptation interfaces studied are (the first five are the main
methods, with control-token tuning as the baseline condition):

\begin{itemize}
\tightlist
\item
  \textbf{LoRA}, which learns low-rank updates in selected transformer
  projections (Hu et al. 2022).
\item
  \textbf{IA3}, which learns multiplicative activation scaling.
\item
  \textbf{BitFit}, which updates bias parameters.
\item
  \textbf{Prefix tuning}, which adds learned virtual tokens (Li and
  Liang 2021).
\item
  \textbf{Full fine-tuning}, which updates all model parameters.
\item
  \textbf{Control-token tuning} (baseline), which trains only the token
  embedding table and the output projection over the frozen transformer
  blocks (Section 5.2), without an adapter.
\end{itemize}

\begin{center}\rule{0.5\linewidth}{0.5pt}\end{center}

\subsection{3. Related Work}\label{related-work}

Chord modeling has a long history in symbolic music research, including
grammar-based and probabilistic approaches to harmonic structure
(Steedman 1984; Rohrmeier 2011; Paiement et al. 2005). Neural chord
generation and harmonization systems have appeared in chord substitution
tools, Bach chorale models, and transformer-based chord systems (Huang
et al. 2016; Liang et al. 2017; Hadjeres et al. 2017; Makris et al.
2020). The Chordinator is especially close in spirit because it studies
style-conditioned chord progression generation across multiple genres
(Dalmazzo et al. 2024). The present report differs by freezing a
single base checkpoint, comparing multiple adaptation interfaces
against a
control-token condition, rotating matched and wrong-genre adapters,
sweeping LoRA rank, and using diagnostics to map representation
boundaries rather than only reporting aggregate style-conditioned
generation accuracy. A stronger archival version should also include
explicit non-neural lower bounds, such as genre-conditioned n-gram or
Markov chord models, to separate neural adaptation gains from what is
already captured by local transition statistics.

Recent controllable-generation work in the ISMIR community also
clarifies the positioning. MusiConGen studies rhythm and chord control
for transformer-based text-to-music generation (Lan et al. 2024),
Content-Based Controls studies controllable music large language
modeling with lightweight tuning (Lin et al. 2024), and MMT-BERT uses
chord-aware symbolic representation for multitrack generation (Zhu et
al. 2024). Those works reinforce that chord and control layers are
important in modern music generation, but they usually aim to improve a
generation system. This report instead asks a boundary question: after
reducing music to chord symbols, which parts of genre-local behavior
remain available to a frozen model and which parts require other musical
layers?

Recent ISMIR evaluation and dataset papers also suggest what the next
version needs. Listener-centered work such as ``Between the AI and Me''
combines quantitative ratings with qualitative reflection for AI- and
human-composed music (Sarmento et al. 2024), and text-to-music
evaluation work has explicitly targeted alignment with human preferences
(Huang et al. 2025). Dataset hygiene and research openness have also
become visible themes through de-duplication work on Lakh MIDI (Choi et
al. 2025) and openness frameworks for music-generative AI (Batlle-Roca
et al. 2025). These directions motivate the conservative claims in this
report: automatic chord prediction is useful evidence, but a stronger
future version needs perceptual evaluation, stricter leakage checks, and
reproducible artifacts.

The base model follows the Music Transformer family (Huang et al. 2019),
but the task is smaller than full polyphonic note modeling: the event
vocabulary and sequence length are both reduced by working at the
chord-symbol level. The previous report in this project studied pop-jazz
rehearsal mixing and produced the F1 fine-tuning run whose released
checkpoint is frozen and used here as the base for multi-genre
adaptation
(\href{https://arxiv.org/abs/2605.04998}{arXiv:2605.04998}). As
verified in v2, that released checkpoint is weight-identical to the
Phase-0 pop baseline (Section 4.1).

Adapter and PEFT methods have become standard in language and music
modeling (Hu et al. 2022; Houlsby et al. 2019; Li and Liang 2021; Han et
al. 2024). In audio-domain music generation, LoRA-style adaptation has
recently been applied to diffusion-based systems such as AudioLDM (Kim
et al. 2025). That line of work targets audio rendering and timbral,
rhythmic, and emotional control. This report instead isolates
chord-symbol sequences as the object of study and asks what genre
information remains after rhythm, timbre, voicing, lyrics, and
performance are removed.

\begin{center}\rule{0.5\linewidth}{0.5pt}\end{center}

\subsection{4. Data and Representation}\label{data-and-representation}

\subsubsection{4.1 Corpora and Genres}\label{corpora-and-genres}

The target genres are blues, bossa nova, Bach chorales, country,
electronic, folk, funk, gospel, hip-hop, R\&B/soul, and rock. All ten
non-chorale genres are drawn from Chordonomicon-derived
chord-transcription splits (Kantarelis et al. 2024), selected by
Chordonomicon's genre tags. The bossa nova slice pools songs tagged
bossa, samba, latin, salsa, or cumbia, so it is a Latin-tagged pool
rather than bossa nova alone. Bach chorales are
treated as a separate tonal-chorale corpus rather than as a typical
contemporary genre. This distinction matters because Bach chorales
behave as a strong outlier in several results.

The base checkpoint is the released F1 artifact from the previous
pop-jazz rehearsal-mix study
(\href{https://arxiv.org/abs/2605.04998}{arXiv:2605.04998}). The F1
fine-tune mixed approximately 1,513 jazz sequences with 10,000
sub-sampled pop sequences (about 87\% pop, 13\% jazz). F1 was chosen as
a design decision rather than an accuracy optimization: the Phase-0
pop base was judged to produce comparatively monotonous progressions, and
the jazz rehearsal mix was intended to retain a richer harmonic
vocabulary (extensions, secondary dominants, ii-V motion), the
output character envisioned for the deployed composition tool. That
design intent stands, but v2 of this report documents that the
released artifact did not embody it: a full SHA-256 comparison over
all 25,841,152 parameters shows the released F1 checkpoint to be
weight-identical to the Phase-0 pop baseline. The F1 run resumed with the Phase-0
optimizer and scheduler state, whose learning rate had decayed to 0, so
its first epoch (epoch 3) updated no weights, and the configured
learning rate and warmup never applied. From epoch 4 it trained under
its cosine schedule from the restored base rate of $3 \times 10^{-4}$
($2.56 \times 10^{-4}$ at epoch 4, down to $4.39 \times 10^{-5}$ at
epoch 8), and its training log shows held-out jazz top-1 rising from
72.86 to a peak of 81.03 at epoch 6, but best-checkpoint selection
minimized validation loss on a pop-dominated validation mix, which rose
monotonically during jazz fine-tuning (0.5578 $\rightarrow$ 0.5901), so
the unchanged epoch-3 checkpoint, the Phase-0 weights, was retained as
``best''. The genuinely
fine-tuned epoch checkpoints were never released. The v1 claim that
the served base retains a richer harmonic vocabulary than the Phase-0
pop base is therefore withdrawn: identical weights imply identical
sampling distributions, and the base-checkpoint comparison that v1
read as ``comparable accuracy between the two bases'' is retroactively
a same-weights control (Section 6.9). The frozen base used throughout
this report is thus, in effect, the Phase-0 model. Its training split
is pop-dominated rather than pop-only: Chordonomicon songs enter it with
a marker mapped from their main genre tag (rock, alternative, punk, and
metal become rock, jazz stays jazz, and every other tag, country, soul,
rap, and electronic included, as well as a missing tag becomes pop), so
of its 544,272 songs, 425,274 carry \texttt{{[}GENRE:pop{]}} (715 of
them from Billboard), 113,389 \texttt{{[}GENRE:rock{]}}, and 5,609
\texttt{{[}GENRE:jazz{]}}. This keeps the base close to the
predominantly pop-adjacent target genres, and it makes rock the one
target whose marker the base had already learned. Because the ten
non-chorale target subsets were drawn later from the same Chordonomicon
file and split on their own, 79.7 to 82.8\% of each of their test sets
also appears, with the same chord sequence, in this Phase-0 training
split. Held-out in this report therefore means held out from the
adapter's own training split, not from the frozen base. The
target set
includes rock, country, folk, R\&B/soul, hip-hop, electronic, gospel,
funk, and blues, with bossa nova as a jazz-adjacent target and Bach
chorales as an outlier.

Table 1 summarizes the per-genre training corpus. Sizes span more than
two orders of magnitude (296 Bach-chorale sequences to 49,388 country
sequences), which motivates the controlled-data-size comparison in
Section 6.10. Bach chorales also have by far the smallest chord
vocabulary (55 unique chords), consistent with their outlier behavior,
while R\&B/soul and funk show the highest chord entropy and the lowest
top-10 coverage.

{\def\LTcaptype{none} 
\begin{longtable}[]{@{}
  >{\raggedright\arraybackslash}p{(\linewidth - 10\tabcolsep) * \real{0.1304}}
  >{\raggedleft\arraybackslash}p{(\linewidth - 10\tabcolsep) * \real{0.1739}}
  >{\raggedleft\arraybackslash}p{(\linewidth - 10\tabcolsep) * \real{0.1739}}
  >{\raggedleft\arraybackslash}p{(\linewidth - 10\tabcolsep) * \real{0.1739}}
  >{\raggedleft\arraybackslash}p{(\linewidth - 10\tabcolsep) * \real{0.1739}}
  >{\raggedleft\arraybackslash}p{(\linewidth - 10\tabcolsep) * \real{0.1739}}@{}}
\toprule\noalign{}
\begin{minipage}[b]{\linewidth}\raggedright
Genre
\end{minipage} & \begin{minipage}[b]{\linewidth}\raggedleft
Train seqs
\end{minipage} & \begin{minipage}[b]{\linewidth}\raggedleft
Mean len
\end{minipage} & \begin{minipage}[b]{\linewidth}\raggedleft
Unique chords
\end{minipage} & \begin{minipage}[b]{\linewidth}\raggedleft
Entropy (bits)
\end{minipage} & \begin{minipage}[b]{\linewidth}\raggedleft
Top-10 cov.
\end{minipage} \\
\midrule\noalign{}
\endhead
\bottomrule\noalign{}
\endlastfoot
blues & 7,955 & 71.0 & 232 & 4.68 & 0.71 \\
bossa nova & 11,452 & 57.8 & 231 & 4.76 & 0.68 \\
Bach chorale & 296 & 31.6 & 55 & 4.52 & 0.70 \\
country & 49,388 & 72.3 & 234 & 4.12 & 0.82 \\
electronic & 12,156 & 81.3 & 230 & 4.66 & 0.69 \\
folk & 48,601 & 77.3 & 237 & 4.38 & 0.78 \\
funk & 2,269 & 84.9 & 223 & 5.27 & 0.58 \\
gospel & 2,997 & 66.3 & 219 & 4.57 & 0.73 \\
hip-hop & 11,216 & 77.1 & 227 & 4.74 & 0.67 \\
R\&B/soul & 7,640 & 78.0 & 239 & 5.46 & 0.54 \\
rock & 39,128 & 79.4 & 240 & 4.55 & 0.74 \\
\end{longtable}
}

\subsubsection{4.2 Tokenization}\label{tokenization}

The tokenizer uses a chord-symbol vocabulary with key, time, bar,
structural, and genre tokens, and every sequence carries its genre
marker in its header, after the key and time-signature tokens. The adaptation runs extend the base vocabulary from 351 to 359
tokens with eight new genre markers (country, R\&B/soul, hip-hop,
electronic, funk, folk, gospel, and Bach chorale). Rock, blues, and
bossa nova use markers already in the base vocabulary, and of these
only rock's was seen in Phase-0 training. Every method trains the full
token embedding table and output projection along with its own
parameters (Section 5.2), so each can learn or adjust its genre
marker.

Chord notation is normalized before tokenization. Enharmonic variants
and quality aliases are mapped into a canonical symbolic space. The goal
is not to model every possible spelling distinction but to create a
stable time-series representation for next-chord prediction and
generation.

\subsubsection{4.3 Data Validity and Repetition
Caveat}\label{data-validity-and-repetition-caveat}

Chord-progression corpora are highly repetitive. Exact duplicate rates
across train and test are usually low, but 4-gram near-overlap is very
high. In the current diagnostic, the mean train-to-test 4-gram overlap
rate is 0.975 across the 11 target genres, with a minimum of 0.933 and a
maximum of 0.998. On average 2.16\% of a test set's songs have the same
chord sequence in their own subset's training split (0.54\% outside
Bach chorales), and 3.63\% once keys are normalized (1.89\%), with the
Bach chorales reaching 18.4\% (7 of 38) and 21.1\% (8 of 38). Earlier
versions gave 0.28\% and 0.46\% (Bach chorales 2.46\% and 2.82\%), the
shares of unique training sequences that recur in test, which are
smaller because each training split is about eight times its test
split. Without the header targets the chorale adapter gains 12.65 pp
over F1 on the 31 test chorales without a training duplicate, against
13.14 pp on all 38. The frozen base saw more:
79.7 to 82.8\% of each non-chorale target test set is in its Phase-0
training split (Section 4.1). Across target subsets the overlap is also
large in places (36.7\% of the country test set is in the folk training
split, and 35.9\% of the folk test set in the country one), which
affects only wrong-genre cells.

This matters for interpretation. The report does not claim open-ended
generalization to completely unseen musical idioms. The main claim is
about held-out chord-transcription distributions under a repetitive
symbolic corpus. The duplicate and near-duplicate diagnostics are
reported to prevent over-reading top-1 gains. The next robustness step
is a strict ``novel-progression'' subset that keeps only test examples
whose key-normalized 4-grams are absent from training, or a
progression-family split that removes songs above a train-test
similarity threshold. If adaptation gains survive there, the result
would directly answer the memorization objection. If they shrink, the
boundary claim becomes sharper.

\subsubsection{4.4 Licensing and
Provenance}\label{licensing-and-provenance}

The contemporary-genre slices are derived from the Chordonomicon dataset
(Kantarelis et al. 2024), which is distributed under a Creative Commons
Attribution-NonCommercial (CC BY-NC 4.0) license. They are used here for
non-commercial research only, and the raw dataset is not redistributed:
released artifacts are limited to trained model weights and the code
that rebuilds the splits and runs the evaluation. Bach chorales are
obtained from the public music21 corpus and are treated as a
tonal-chorale reference rather than a contemporary genre. Because public
entry points for Chordonomicon have shown differing license labels over
time, I record the source used here: the Hugging Face dataset entry
\texttt{ailsntua/Chordonomicon}, labeled CC BY-NC 4.0 at download, whose
CSV file (264,198,044 bytes) has SHA-256
\texttt{9d2f4ccdc876a4e816712f128e4772b2af558c2a2923cce5be3a0364fbad9a8d}.
The upstream revision string was not recorded, so this checksum is the
content pin.

\begin{center}\rule{0.5\linewidth}{0.5pt}\end{center}

\subsection{5. Method}\label{method}

\subsubsection{5.1 Base Model}\label{base-model}

The frozen base is a 25.7M-parameter Music Transformer checkpoint from
the previous pop-jazz study, released as F1 but verified
weight-identical to the Phase-0 pop baseline by a full SHA-256
comparison over all 25,841,152 parameters (Section 4.1). The
architecture uses relative-position attention with a chord-symbol
vocabulary and sequence length appropriate for chord progressions. All
target-genre adaptation experiments reported in the main grid begin
from this same released base: every per-genre adapter was trained on,
and is self-consistent with, these weights, and all released
evaluation numbers were produced with these served artifacts and
reproduce on independent CPU hardware to within 0.02 pp at the model
level.\footnote{Training used fp16 mixed precision. The residual
cross-hardware difference is the GPU mixed-precision versus CPU
full-precision numeric gap, well below the three-seed spread of the method macros (0.09 to 0.32
pp, Section~6.9).}

\subsubsection{5.2 Adaptation Methods}\label{adaptation-methods}

For each target genre, I train LoRA, IA3, BitFit, prefix tuning, and
full fine-tuning under a uniform 8-epoch setting and three random seeds.
LoRA ranks are selected by a prior validation rank sweep over ranks 4,
8, 16, 32, and 64. Prefix tuning uses 20 virtual tokens. For PEFT methods, trainable parameters include the method-specific
parameters plus the full token embedding table and output projection
(367,616 parameters, two tables of $359 \times 512$), which every method
unfreezes so that the new genre markers can be learned.
The methods differ sharply in trainable footprint:

{\def\LTcaptype{none} 
\begin{longtable}[]{@{}lrr@{}}
\toprule\noalign{}
Method & Trainable params & \% of model \\
\midrule\noalign{}
\endhead
\bottomrule\noalign{}
\endlastfoot
IA3 & 375,808 & 1.5\% \\
BitFit & 413,184 & 1.6\% \\
Prefix tuning & 531,456 & 2.1\% \\
LoRA (per-genre rank 4--64) & 465,920 -- 1,940,480 & 1.8--7.5\% \\
Full fine-tuning & 25,849,344 & 100\% \\
\end{longtable}
}

The four parameter-efficient methods land within roughly half a
percentage point of full fine-tuning on macro top-1 (Section 6.1),
with most matching or exceeding it, while training only 1.5 to 7.5\% as many parameters, so the modularity argument for adapters carries little
accuracy cost.

The control-token baseline freezes every transformer block and trains
only the token embedding table and the output projection (the same
367,616 parameters that the other methods also unfreeze, 1.4\% of the
model), so the genre marker and the chord-token embeddings move while
attention and feed-forward weights stay fixed. It runs over the same
target genres and seeds and tests whether adapter gains need capacity
inside the transformer blocks or whether much of the effect is
reachable through the input and output layers alone.

\subsubsection{5.3 Evaluations}\label{evaluations}

The main metric is held-out next-token top-1, with top-5 and loss
retained as supporting metrics. It is computed over every non-padding
target token, so bar markers, the end token, and the three header tokens
that open every sequence (key, time signature, and genre marker) count
alongside chords. A model predicts each header token from the header
tokens before it, so it makes nearly the same header prediction for
every test sequence, and each header target (0.79 to 1.85\% of a
genre's test targets) moves a whole run at once. F1 never predicts a
target genre's marker, while an adapter trained on that genre predicts
it every time. Chordonomicon has no key annotation, so every song of the
ten non-chorale test sets carries the placeholder
\texttt{{[}KEY:Cmaj{]}} and the 4/4 time signature, which every model
predicts. Twelve-key augmentation leaves a model's key prediction after
\texttt{{[}BOS{]}} close to uniform (F1 gives B$\flat$ major 0.093 and C
major 0.087), so which major key a run predicts there is close to
arbitrary, and a run that predicts C major scores one more target per
sequence on every non-chorale test set. F1 does not, and of the eleven
released adapters the blues, bossa nova, country, and gospel ones do
(measured for all of them on all 11 test sets). The runs whose
checkpoints were not kept cannot be checked directly, but within each
genre their scores fall on levels one such step apart, agreeing within
about 0.15 pp inside a level. Taking the marker to be predicted, as
every released adapter predicts it, I read the upper level as the runs
that predicted C major, and the ten released non-chorale adapters agree
with this reading. The nine BitFit cells in country, folk, and rock
fall on neither level (0.9 to 1.8 pp below the lower one, Section 5.4),
so their key is assumed, not read, to be other than C major. On the
Bach-chorale test set, whose songs carry real keys, F1 and the
non-chorale adapters also miss 20 of the 38 time-signature targets,
which the chorale adapter gets right. Sections 6.1, 6.3, 6.4, and 6.10
therefore also report results scored without the three header targets
(header-free). The main PEFT grid is evaluated over 11 genres, 5
methods, and 3 seeds. Method comparisons are analyzed with Wilcoxon
signed-rank tests over per-genre seed means, followed by
Holm-Bonferroni and Benjamini-Hochberg correction.

F1 is also handicapped by the marker as an input. Phase 0 maps every
Chordonomicon tag onto pop, rock, or jazz (Section 4.1), so the blues
and bossa nova markers of the base vocabulary never occur as inputs in
its training, and the eight new markers start as copies of the equally
unused \texttt{{[}GENRE:none{]}} row. Rock's marker is the only target
marker that F1 has read in training. To measure the cost, I rescored F1
on CPU on every test set with the input marker replaced by
\texttt{{[}GENRE:pop{]}} and compared it with the released matched LoRA
adapters, all header-free. With the pop marker F1 gains 1.0 to 1.9 pp
in the nine non-chorale genres other than rock, where the swap changes
nothing, and 3.6 pp on Bach chorales, and it comes within 0.12 pp of the
released matched adapter in every non-chorale genre (Section 6.1).
\texttt{{[}GENRE:rock{]}} in place of the pop marker gives the same
within about 0.1 pp on the five smallest test sets, and on the 787 test songs of blues, bossa nova, funk,
gospel, and R\&B/soul whose chord sequence is not in the Phase-0
training split, the matched adapters lead F1 given the pop marker by
0.03 pp pooled.

Additional diagnostics include:

\begin{itemize}
\tightlist
\item
  LoRA selected-rank multi-seed reliability.
\item
  LoRA rank sweep with Wilson confidence intervals.
\item
  Wrong-genre adapter rotation.
\item
  Chord-only genre classification using chord tokens only.
\item
  Generated-output statistics comparing F1 and F1+LoRA.
\item
  Real-song chord-chart evaluation.
\item
  Duplicate and near-duplicate diagnostics.
\end{itemize}

\subsubsection{5.4 Reproducibility}\label{reproducibility}

The frozen base checkpoint and all per-genre adapters are released on
Hugging Face (huggingface.co/PearlLeeStudio), and the code, including
split creation, the training configurations, and the evaluation
scripts, is public at github.com/PearlLeeStudio/TheArtist. The derived
split files are not distributed. Because genre tags are assigned first
match within one run of \texttt{extract\_genre\_subsets.py}, the genre
subsets reproduce only when it is run twice with \texttt{-{}-merge},
first for \texttt{country} and then for \texttt{funk gospel blues\_extra
rnb\_soul hip\_hop bossa electronic folk rock\_extra} in that order. The
Bach-chorale subset comes from \texttt{extract\_bach\_chorales.py}. The complete 165-cell grid
and all diagnostics in this report were trained and evaluated on a
single consumer laptop GPU (NVIDIA GeForce RTX~4070 Laptop, 8~GB), in
fp16 mixed precision, which indicates that systematic per-genre
chord-symbol adaptation studies are feasible without dedicated training
infrastructure. The selection-corrected jazz retrains and their
multi-seed reproduction (Section~6.9) were instead trained on cloud
Tesla~T4 GPUs (Colab and Kaggle), also in fp16 mixed precision. Their
three-seed numbers come from the training-log evaluations. The Phase-0
baseline and the released retrain were also re-evaluated in full
precision on CPU on the same test set, and the Phase-0 CPU value agrees
with its GPU value to within 0.02~pp.

One artifact-selection detail matters when reproducing the LoRA
numbers. The per-genre LoRA ranks used for the selected-rank
evaluation in this report were chosen by highest validation top-1 over
the rank sweep. The released adapters on Hugging Face were instead
selected, deliberately, by lowest validation loss with a top-1
tiebreak, consistent with training-time best-checkpoint selection
throughout the pipeline (\texttt{model/\_release\_best\_ranks.py}).
The two criteria pick the same rank for five genres but differ for
six: electronic (selected 8 vs released 4), folk (32 vs 4), funk (64
vs 8), hip-hop (4 vs 32), R\&B/soul (4 vs 8), and rock (16 vs 8).
Readers reproducing table numbers from the released adapters should
therefore expect the released-rank cells of the rank-sweep table
(Section 6.3), not the selected-rank row.

One evaluation detail affects the full fine-tuning and BitFit cells.
Their test evaluation rebuilt each model with tied input and output
embeddings and then loaded the separately trained tables, so the
trained output projection also served as the input embedding.
Reproducing that substitution on released LoRA adapters, which train
both tables at the BitFit learning rate, changes test top-1 by at most
0.05 pp for the funk, gospel, blues, and rock adapters, but it moves the
country adapter's key prediction off C major, a whole key step (0.92 pp,
Section 5.3), and the two tables differ most for the adapters trained on
the largest corpora (relative difference 0.40 to 0.62 for rock, folk,
and country, against 0.19 to 0.28 for funk, gospel, and blues). Full
fine-tuning used a 30 times lower rate, so I expect its cells to be
little affected. BitFit's header-free gains in country, folk, and rock,
the three largest corpora, sit 1.2 to 1.9 pp below those of the other
methods while matching them in the other non-chorale genres and at a
matched data size (Sections 6.1 and 6.10). BitFit's own training logs
point to the reload. On the validation split its seed-42 runs in these
three genres end level with IA3 (loss 0.4976, 0.5195, and 0.5717
against 0.4986, 0.5193, and 0.5719), while the reloaded test loss of
the same runs is 0.6245, 0.5479, and 0.7218, against 0.4939, 0.5256,
and 0.5705 for IA3. The deficit therefore most likely comes from the
reload and not from BitFit itself, although no BitFit checkpoint was
kept to confirm it.

\begin{center}\rule{0.5\linewidth}{0.5pt}\end{center}

\subsection{6. Results}\label{results}

\subsubsection{6.1 Main 165-Cell Adaptation
Grid}\label{main-165-cell-adaptation-grid}

The complete main grid contains 5 methods $\times$ 11 genres $\times$ 3 seeds = 165
cells. All five methods improve over the frozen F1 base on macro
held-out top-1.

{\def\LTcaptype{none} 
\begin{longtable}[]{@{}
  >{\raggedright\arraybackslash}p{(\linewidth - 10\tabcolsep) * \real{0.1304}}
  >{\raggedleft\arraybackslash}p{(\linewidth - 10\tabcolsep) * \real{0.1739}}
  >{\raggedleft\arraybackslash}p{(\linewidth - 10\tabcolsep) * \real{0.1739}}
  >{\raggedleft\arraybackslash}p{(\linewidth - 10\tabcolsep) * \real{0.1739}}
  >{\raggedleft\arraybackslash}p{(\linewidth - 10\tabcolsep) * \real{0.1739}}
  >{\raggedleft\arraybackslash}p{(\linewidth - 10\tabcolsep) * \real{0.1739}}@{}}
\toprule\noalign{}
\begin{minipage}[b]{\linewidth}\raggedright
Method
\end{minipage} & \begin{minipage}[b]{\linewidth}\raggedleft
Macro top-1
\end{minipage} & \begin{minipage}[b]{\linewidth}\raggedleft
Macro Delta vs F1
\end{minipage} & \begin{minipage}[b]{\linewidth}\raggedleft
Non-chorale Delta
\end{minipage} & \begin{minipage}[b]{\linewidth}\raggedleft
Beats F1
\end{minipage} & \begin{minipage}[b]{\linewidth}\raggedleft
Best genres
\end{minipage} \\
\midrule\noalign{}
\endhead
\bottomrule\noalign{}
\endlastfoot
LoRA & 82.51 & +3.61 pp & +2.41 pp & 11/11 & 4/11 \\
IA3 & 82.41 & +3.51 pp & +2.55 pp & 11/11 & 4/11 \\
Prefix tuning & 82.23 & +3.33 pp & +2.49 pp & 11/11 & 2/11 \\
Full fine-tuning & 81.97 & +3.07 pp & +2.38 pp & 11/11 & 0/11 \\
BitFit & 81.79 & +2.89 pp & +1.97 pp & 10/11 & 1/11 \\
\end{longtable}
}

The headline is not that LoRA wins. LoRA and IA3 are strongest by macro
top-1 and genre win count, but the pairwise significance table does not
support a decisive winner after Holm or Benjamini-Hochberg correction.
The method-level result is better read as a robust adaptation effect
across several interfaces.

Part of every gain is the header targets (Section 5.3). Without them,
the macro gains are +2.38 (LoRA), +2.16 (IA3), +2.02 (prefix tuning),
+1.88 (full fine-tuning), and +1.64 pp (BitFit), and outside Bach
chorales LoRA, IA3, prefix tuning, and full fine-tuning give the same
gain, +1.30 to +1.34 pp, while BitFit gives +0.85 pp. Within each
non-chorale genre the five methods' header-free gains lie within
0.13 pp of each other, apart from BitFit in country, folk, and rock, the
three largest corpora, where it sits 1.2 to 1.9 pp below the others,
most likely because of how its test cells were reloaded (Section 5.4).
Outside Bach chorale and those BitFit cells, the method differences in
the table above, its genre win counts included, and the per-genre
winners below therefore come from the header targets, mostly from how
many of a method's runs predicted the C-major key. The 11-genre macro
keeps the same method order without the header targets, because those
two sources of difference remain, and LoRA's lead there comes from Bach
chorale. For the runs without kept checkpoints, the marker is taken to
be predicted as by the released adapters, and the key is read from the
score level (Section 5.3).

The non-chorale header-free gain is itself about the size of F1's
marker handicap (Section 5.3). Against F1 given the pop marker, the
released matched LoRA adapters gain $-0.02$ to +0.12 pp in the ten
non-chorale genres (+0.06 pp on average) and 9.6 pp on Bach
chorales, and set against the same baseline the main-grid gains of
LoRA, IA3, prefix tuning, and full fine-tuning outside Bach chorales
become +0.05 to +0.10 pp. Outside Bach chorales the measurable benefit of
adaptation on these test sets is therefore a genre marker that the
model can use rather than measurably better chord prediction, measured
directly for the released LoRA adapters and through the shifted
baseline for the other methods. Rock gains about nothing without the header
targets (+0.01 to +0.06 pp for the methods other than BitFit). The
overlap with Phase-0 training does not explain this, since 79.7 to
82.8\% of every non-chorale test set is in that split (Section 4.1).
Rock is the one target whose marker F1 has read in training, so it is
the one genre in which F1 starts without the handicap.

The best method varies by genre:

{\def\LTcaptype{none} 
\begin{longtable}[]{@{}llrr@{}}
\toprule\noalign{}
Genre & Best method & Best top-1 & Delta vs F1 \\
\midrule\noalign{}
\endhead
\bottomrule\noalign{}
\endlastfoot
blues & IA3 & 84.43 & +3.06 pp \\
bossa nova & LoRA & 82.37 & +3.55 pp \\
Bach chorale & LoRA & 60.38 & +15.54 pp \\
country & LoRA & 85.78 & +3.22 pp \\
electronic & IA3 & 87.02 & +2.44 pp \\
folk & IA3 & 85.41 & +2.86 pp \\
funk & IA3 & 84.59 & +2.41 pp \\
gospel & LoRA & 82.12 & +3.23 pp \\
hip-hop & BitFit & 88.91 & +2.46 pp \\
R\&B/soul & Prefix tuning & 85.33 & +2.59 pp \\
rock & Prefix tuning & 84.65 & +1.72 pp \\
\end{longtable}
}

On the header-free reading above, these per-genre winners do not rank
the methods outside Bach chorale. Bach chorale is the clearest outlier.
Excluding it, gains are still positive but more modest. This supports the boundary framing:
chord-symbol adaptation is useful, but much of the non-chorale genre
range lies in overlapping harmonic territory.

\begin{figure}
\centering
\includegraphics[width=0.85\linewidth,height=\textheight,keepaspectratio]{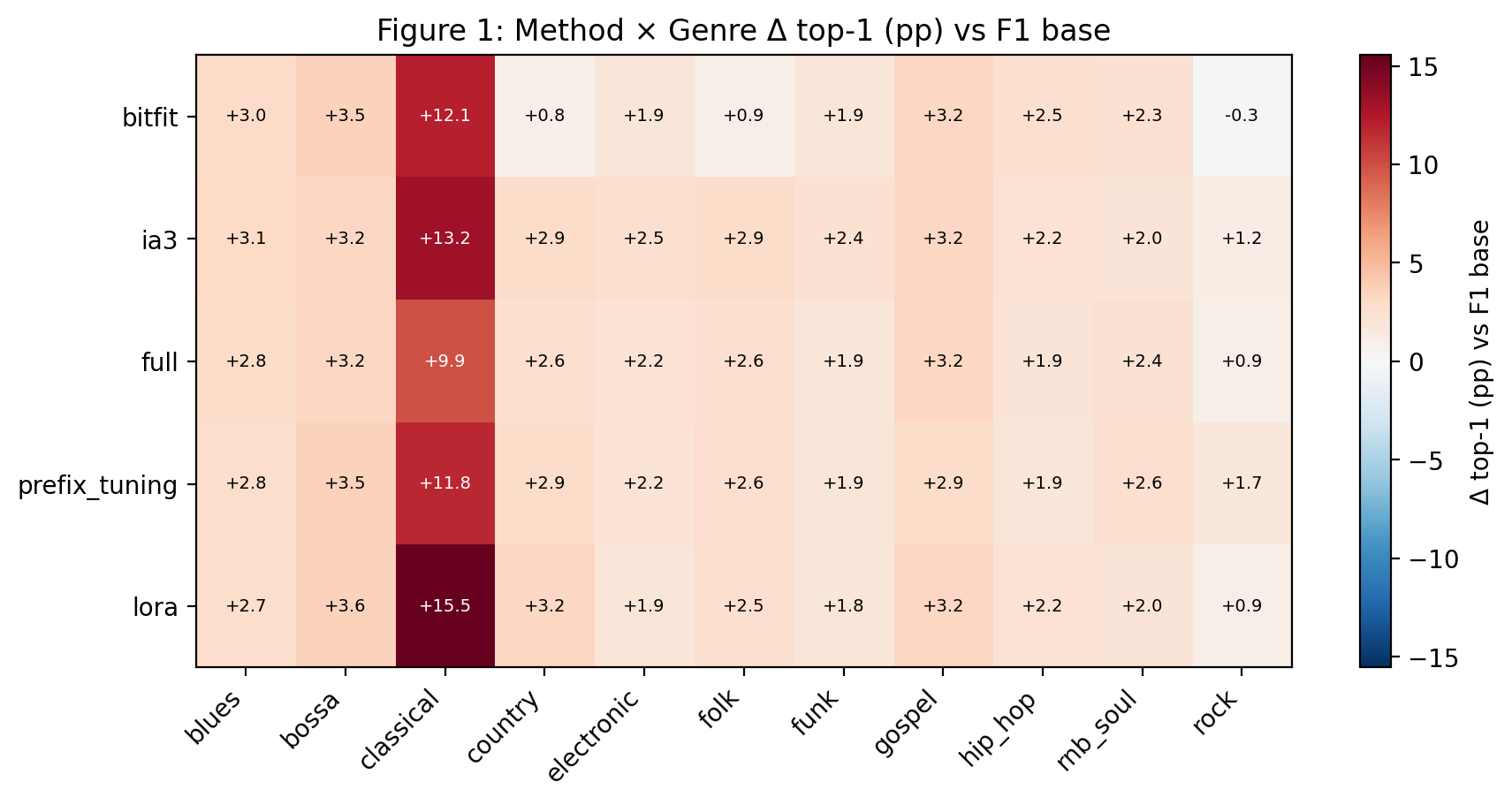}
\caption{Method $\times$ genre Delta top-1 (pp) versus the frozen F1
base, all targets scored. Outside Bach chorale the differences between
methods are mostly key-token steps, BitFit's low country, folk, and
rock cells most likely come from a test-time reload (Section 5.4), and
the gains themselves are about the size of F1's marker handicap
(Section 5.3).}
\end{figure}

\subsubsection{6.2 Control-Token Baseline}\label{control-token-baseline}

The control-token baseline is strong. It reaches macro top-1 82.01, a
validation-split value logged at the end of training (token level, 11
genres, three seeds pooled), which is +3.11 pp above F1's test macro
(78.90), with non-chorale Delta +2.26 pp. No control-token checkpoint
was kept, so no test-split score exists for it, whereas F1 and every
method row are test-split scores. On the same validation split F1
scores 78.77, so the like-for-like gap is +3.24 pp, and +2.28 pp
outside Bach chorales. Relative to the control-token baseline, mean
method gaps (test-split method macros minus this validation-split
value) are small:

\begin{center}
\begin{tabular}{@{}lrr@{}}
\toprule
Method & Mean gap vs control-token & Better genres \\
\midrule
LoRA & +0.49 pp & 6/11 \\
IA3 & +0.40 pp & 9/11 \\
Prefix tuning & +0.22 pp & 6/11 \\
Full fine-tuning & -0.04 pp & 5/11 \\
BitFit & -0.22 pp & 5/11 \\
\bottomrule
\end{tabular}
\end{center}

These gaps are all smaller than one key step (Section 5.3) and set
test means against validation values, so, like the counts of Section
6.1, the better-genre counts do not rank the methods. This result
weakens any claim that a particular adapter family is necessary for
genre adaptation. It strengthens a different claim: the frozen chord
model already contains reusable harmonic structure, and small
adaptation interfaces can expose genre-local behavior where the target
harmony departs from the base, as for Bach chorales. The
control-token baseline trains the genre marker's embedding along with
the other token embeddings, which fits the finding of Section 5.3 that
outside Bach chorales a usable marker is most of what adaptation
supplies. Adapters
remain useful for modular serving, per-genre replacement, and isolating
genre-specific parameters, but the accuracy story is not adapter-only.
The reusable insight is therefore that control-token conditioning lands
in the same accuracy band as adapter tuning, not that LoRA narrowly
beats IA3.

\subsubsection{6.3 LoRA Rank Sweep and Selected-Rank
Reliability}\label{lora-rank-sweep-and-selected-rank-reliability}

The selected-rank LoRA evaluation improves over F1 in all 11 target
genres. Across its three-seed summary, the macro top-1 gain is +3.71 pp.
This summary differs from the main-grid LoRA cells of Section 6.1
(macro gain +3.61 pp) in four of its 33 seed cells, each by one key
step (Section 5.3), and agrees with the others within 0.05 pp.
Over the 11 per-genre mean deltas the median is +2.74 pp and the
non-chorale mean is +2.52 pp.~The minimum per-genre mean gain is rock at
+1.14 pp, and the maximum is Bach chorale at +15.53 pp.

Rank scaling is shallow for most genres. The selected ranks (rank 4
for country, gospel, hip-hop, and R\&B/soul, rank 32 for blues and folk,
rank 64 for Bach chorale and funk) are mostly not preferences: in six
genres (blues, bossa nova, electronic, folk, gospel, rock) the selected
rank is the only one whose run predicted the C-major key (Section 5.3),
one step above the other ranks, and apart from that step the ranks of a
non-chorale genre differ by at most 0.14 pp. Only Bach chorale shows a
clear rank trend. This suggests that the main bottleneck
is not simply adapter rank. It is the amount and kind of genre
information available in chord-symbol sequences.

\begin{figure}
\centering
\includegraphics[width=0.85\linewidth,height=\textheight,keepaspectratio]{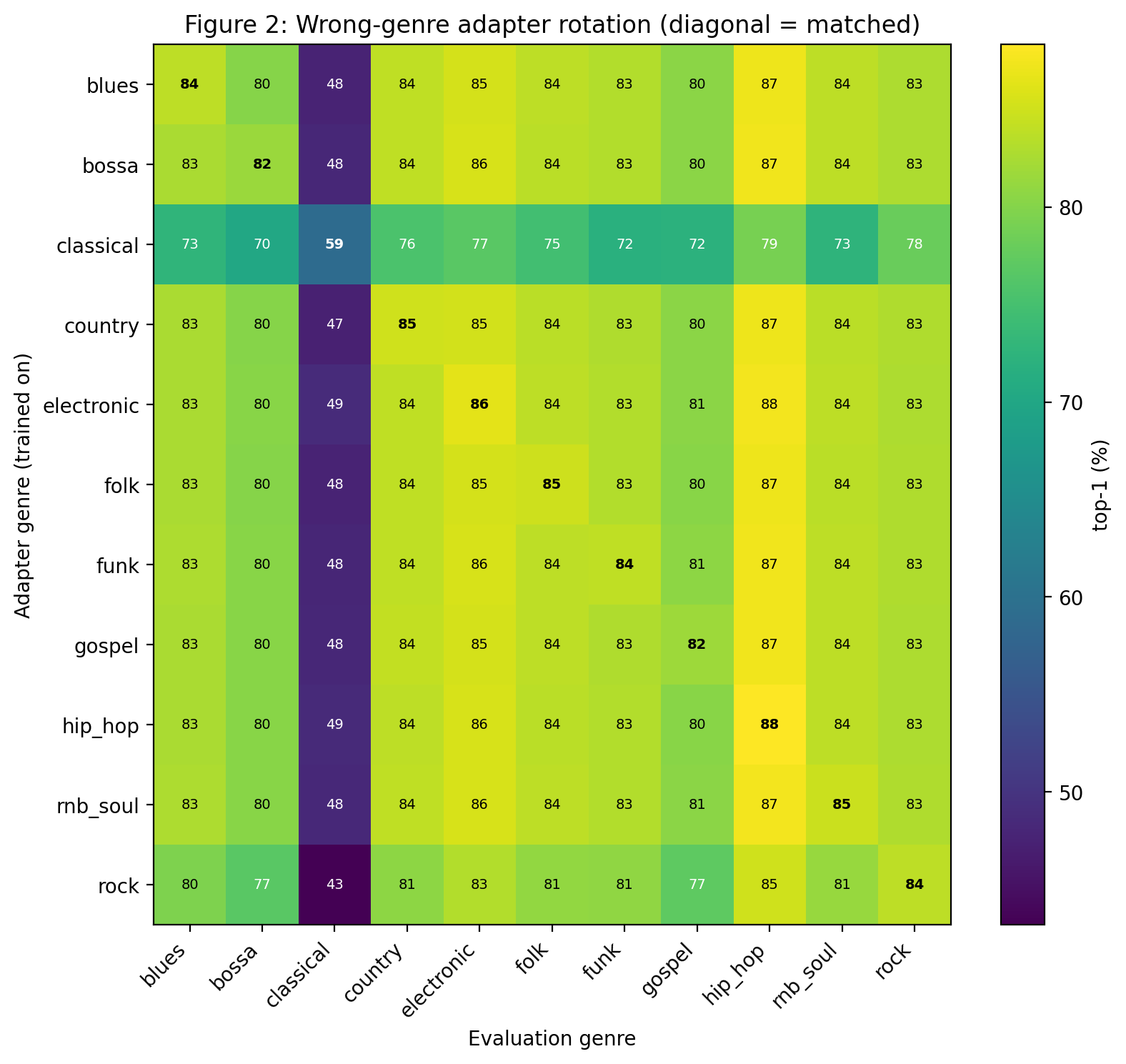}
\caption{Wrong-genre adapter rotation (Section 6.4). The diagonal is the
matched adapter. All targets are scored, so the diagonal includes the
genre marker, which only the matched adapter predicts.}
\end{figure}

\begin{figure}
\centering
\includegraphics[width=0.85\linewidth,height=\textheight,keepaspectratio]{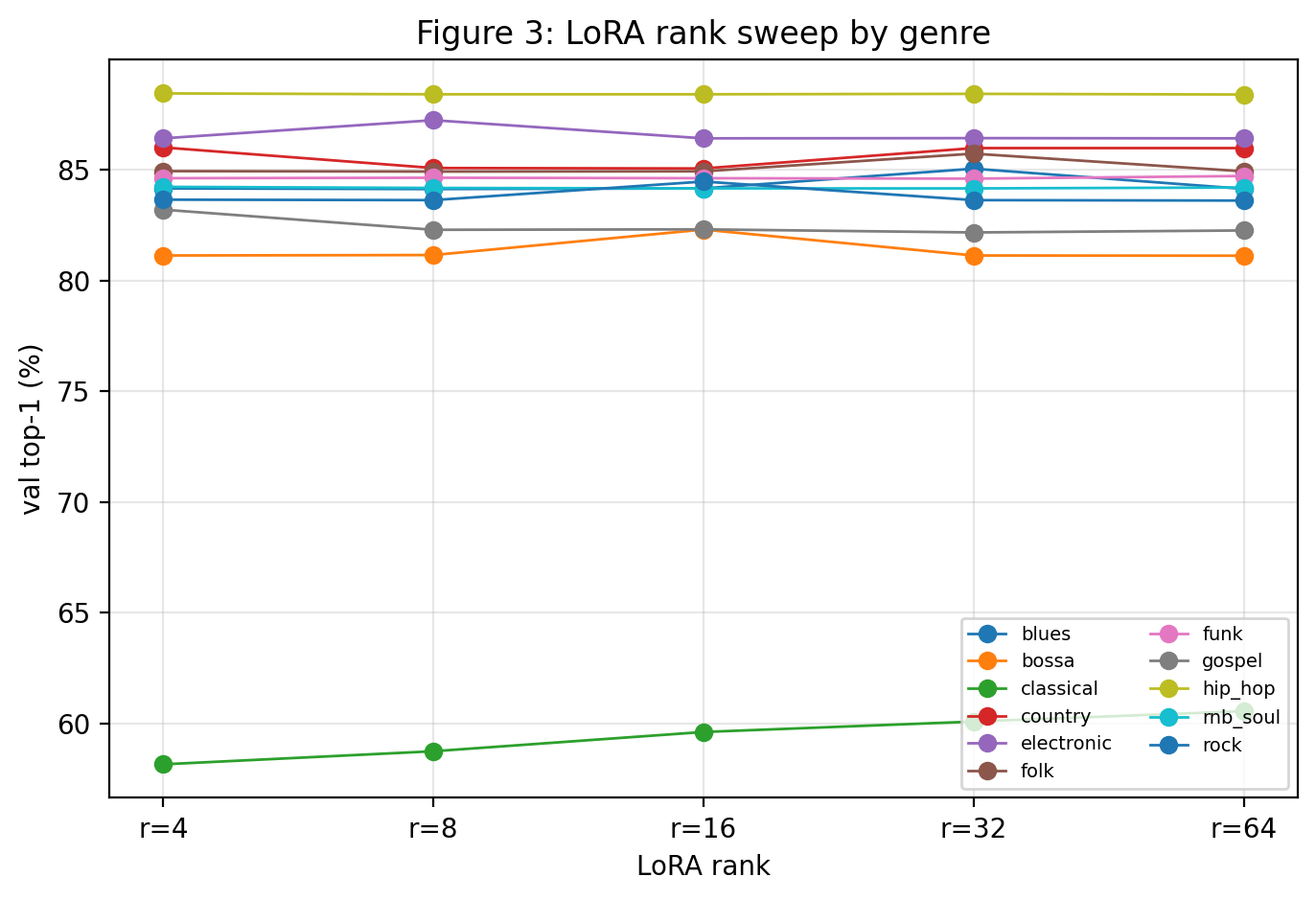}
\caption{LoRA validation top-1 across rank for each genre, all targets
scored. Ranks that sit one step above the others are runs that
predicted the C-major key.}
\end{figure}

\subsubsection{6.4 Wrong-Genre Adapter
Rotation}\label{wrong-genre-adapter-rotation}

The wrong-genre rotation evaluates each adapter on each target genre.
The matched adapter is better than the off-diagonal average in 11/11
eval genres and is at least as strong as the best off-diagonal adapter
in 11/11. The mean matched-minus-off-diagonal-average gap is +3.07 pp.
These counts include the header targets (Section 5.3). Like F1, the
adapters in this matrix predict a key other than C major on the
Chordonomicon test sets (the pop-family adapters other than the matched
one lie within 0.36 pp of each other on every such set, under one key step, and the
rock and chorale adapters are never the best wrong adapter there), so
the header targets that separate them are the genre marker, which only
the matched adapter predicts, and on the Bach-chorale set the time
signature. Without the header targets, the mean gap is +2.06 pp and the
matched adapter still beats the off-diagonal average in 11/11 genres,
but it is at least as strong as the best off-diagonal adapter in 10/11
(hip-hop is 0.01 pp behind), and its margin over that adapter is
$-0.01$ to $+0.20$ pp for the ten non-chorale genres, against 7.6 to
8.3 pp for Bach chorale (10.32 pp with the header targets). The 81
off-diagonal cells that beat F1 do so with or without the header
targets, but they are exactly the nine pop-family adapters other than
rock's, each trained with a marker row that F1 never read, on the nine
test sets other than their own and rock's, where F1 reads an untrained
marker (Section 5.3). On the rock set, whose marker F1 has read in
training, no wrong adapter beats F1, and the rock adapter beats F1 on no
other set. Against F1 given the pop marker, only 4 of the 90
wrong-genre cells on the nine non-chorale sets other than rock score
higher without the header targets, by at most 0.07 pp. The rotation was
run on adapter files that do not all match the released ones: on the
cells checked, the released electronic, funk, and hip-hop adapters
reproduce the matrix within 0.07 pp, the released blues, bossa nova,
gospel, and country adapters, which predict C major, score 0.24 to
1.06 pp above it on the Chordonomicon sets (about one key step for
blues, bossa nova, and gospel, less for country) and $-0.78$ to
$+0.24$ pp off it on the Bach-chorale set, and the released
Bach-chorale adapter scores 1.31 pp
above the matrix diagonal on its own test set, a gap that no header
target explains. A reproduction from the released adapters should
therefore be compared on the header-free margins, and its Bach-chorale
row will not match the matrix.

At the same time, off-diagonal adapters are not weak. In the current
matrix, 81 out of 110 off-diagonal cells exceed the F1 baseline for the
eval genre. This means the wrong-genre rotation should not be interpreted as proof
that adapters are purely genre-specific. It does not show a generic
corpus-adaptation effect either, because the wrong adapters beat F1
only where F1 reads an untrained marker, and F1 given the pop marker
scores as well as nearly every wrong adapter without the header
targets. Outside Bach chorales the matched adapter adds a genre-local
advantage over the average wrong adapter but hardly any over the best
one once the header targets are set aside, so what an adapter supplies
there is mostly a genre marker the model can use. This is the second
reusable finding: a frozen base has to be scored with a genre marker it
was trained on before wrong-genre or matched adapters are credited with
harmonic adaptation.

\subsubsection{6.5 Generated-Output
Statistics}\label{generated-output-statistics}

Generated-output diagnostics compare F1 and F1+LoRA over sampled
continuations. LoRA outputs move closer to the target training
distribution:

{\def\LTcaptype{none} 
\begin{longtable}[]{@{}lrl@{}}
\toprule\noalign{}
Metric & Mean LoRA minus F1 & Direction \\
\midrule\noalign{}
\endhead
\bottomrule\noalign{}
\endlastfoot
Unique chords & -23.64 & lower in 10/11 genres \\
Chord entropy & -0.59 bits & lower in 10/11 genres \\
Repetition rate & -0.119 & lower in 10/11 genres \\
Chord-vocabulary KL vs train & -0.677 & lower in 11/11 genres \\
Bigram KL vs train & -2.709 & lower in 11/11 genres \\
\end{longtable}
}

This supports a distribution-matching interpretation. The adapted model
produces chord sequences whose unigram and bigram distributions are
closer to the target training set. But it also reduces unique chord
count and entropy in most genres, so the result should not be described
as greater creativity or diversity. The safer conclusion is that adaptation improves genre-local
distribution alignment while often narrowing output diversity. Much of
the shift is F1's rather than the adapter's. The F1 samples are
prompted with the target marker, which F1 never read in training, and
on rock, the one target whose marker F1 has read, the LoRA changes these
statistics far less (unique chords +4 against $-6$ to $-37$ in the
other genres, bigram KL $-0.21$ against $-1.45$ to $-6.57$), which is why rock is the
exception in the unique-chord and entropy counts.

\subsubsection{6.6 Chord-Only Genre
Classification}\label{chord-only-genre-classification}

A diagnostic classifier using only chord tokens achieves accuracy 0.247,
balanced accuracy 0.225, and macro F1 0.171, above the 11-class chance
balanced accuracy of 0.091. This confirms that chord sequences carry
genre information. But the low macro F1 also confirms that the signal is
incomplete.

Country is the strongest class by F1, while many genres remain weakly
separable. The correlation between classifier recall and LoRA gain is
unstable once Bach chorale is removed. The classifier therefore supports
the main thesis: chord-symbol time series contain measurable genre
information, but not enough to represent full genre identity.

\subsubsection{6.7 Real-Song Chord-Chart
Evaluation}\label{real-song-chord-chart-evaluation}

The real-song subset is a sanity check rather than primary evidence.
Scored on all targets, target LoRA beats F1 on the per-genre mean in
all 11 target-LoRA genres. The mean target-minus-F1 Delta is +2.52 pp,
the median is +1.36 pp, the smallest gain is electronic at +0.54 pp, and
the largest is Bach chorale at +12.33 pp. These deltas include the
header targets, and the evaluation gives F1 no genre marker at all for
the eight new genres, whose markers are outside its vocabulary, and an
untrained one for blues and bossa nova. Rescored on CPU from the
released adapters without the header targets, the target LoRA still
leads F1 in all 11 genres (median +0.81 pp, partly because F1 without
a marker predicts one at the first bar target), but against F1 given the
pop marker it leads in only 5 of the 11 (median $-0.06$ pp), by at most
0.49 pp outside Bach chorales and by 10.2 pp on them.

On all targets the direction matches the held-out split results, which
is useful. But
the subset has only 10 songs per genre and is biased toward chord-rich
transcriptions. It should be read as model-card evidence, not as a
substitute for a controlled listening or musician evaluation.

\subsubsection{6.8 Dataset Validity}\label{dataset-validity}

Exact duplicates are mostly low outside a few genres, but near-duplicate
4-gram overlap is high across the board. This is not surprising for
chord progressions: common harmonic loops and cadences recur heavily
across songs. The report therefore frames the task as held-out
prediction within chord-transcription distributions, not as
unconstrained generalization to novel musical styles.

\subsubsection{6.9 Base-Checkpoint Ablation (in v2: a Same-Weights
Control)}\label{base-checkpoint-ablation}

This experiment was designed to test whether the adaptation result
depends on the specific F1 base: I re-adapt each genre from the earlier
Phase-0 pop checkpoint (full fine-tuning and selected-rank LoRA, with funk at rank 4 rather
than its selected 64, 8 epochs, seed 42) and compare against adaptation from the
released F1 checkpoint. Macro top-1 is essentially unchanged between the two runs. Compared on
the same validation split, read from the training logs of both runs,
the Phase-0-minus-F1 macro difference is $-0.13$ pp for LoRA and
$+0.02$ pp for full fine-tuning, and eight of the eleven genres agree to
the logged decimal under each method. Two of the six cells that differ
(country under LoRA and blues under full fine-tuning, 0.9 pp each) are
one key step (Section 5.3), and Bach chorale under full fine-tuning
differs by 1.2 pp on its 37-sequence validation split. For seven LoRA
genres (bossa nova, country, electronic, folk, hip-hop, R\&B/soul, rock)
and for Bach chorale under full fine-tuning, the F1-side log in the
record is that of an earlier 5-epoch run of the same cell, not of the
8-epoch rerun behind Section 6.1, so this comparison is indicative.
Versions before v5 reported $-0.38$ and $+0.21$ pp, with per-genre gaps
up to 1.6 pp, because they subtracted F1 test-split scores from Phase-0
validation-split logs.

In v2 this section must be reinterpreted. As documented in Section 4.1,
the released F1 checkpoint is weight-identical to the Phase-0
checkpoint, so the two ``bases'' compared here are the same weights and
the experiment is retroactively a same-weights control rather than a
base ablation. Read that way, it carries two useful pieces of
information. First, re-adapting twice from the same
weights moves the macro by at most 0.13 pp on one split, while the three
seeds of the main grid spread each method's macro over 0.09 to 0.32 pp.
The seed spread is the run-to-run noise against which the method gaps
should be read. Second, the ``essentially unchanged''
outcome independently corroborates the weight identity established by
hashing: re-adapting from the two checkpoints behaves exactly as
re-adapting twice from one. What this section can no longer claim is
base-robustness: whether the adaptation results survive a genuinely
different (e.g., jazz-adapted) base remains untested (Section 8). A
genuinely jazz-adapted base does now exist via the selection-corrected
retrain (released \texttt{ft-pop80-v2}, hash-distinct from Phase-0,
with its jazz-val selection rule reproduced across three matched-data
retrain seeds), so this gap is one of experimental effort
rather than artifact availability. Re-adapting the genre adapters over
that base is still future work. The
episode is also an instance of the phenomenon this project studies:
metric-optimal checkpoint selection on a mismatched validation
distribution can silently discard adaptation. (Single seed,
descriptive.)

\subsubsection{6.10 Controlled-Data-Size
Comparison}\label{controlled-data-size-comparison}

Because corpus sizes vary by more than two orders of magnitude across
genres, the main-grid method ranking could be data-size driven. I
sub-sample ten genres (Bach chorales excluded because their validation
split is too small to sub-sample reliably) to the funk reference size, re-train
LoRA, IA3, BitFit, and full fine-tuning on the matched-size splits, and
evaluate each model on the original held-out test set.

At matched size the macro test top-1 over the ten non-chorale genres
(seed 42) is IA3 85.17, full fine-tuning 85.09, BitFit 84.78, LoRA
84.44. This order is the key token (Section 5.3). In every genre the
four scores fall on two levels whose gap matches one key step (0.79 to
1.15 pp) and agree within 0.07 pp inside a level, and IA3, full
fine-tuning, BitFit, and LoRA sit on the upper level in 8, 7, 4, and 0
of the ten genres. Without the header targets their gains over F1 are
+1.29, +1.30, +1.28, and +1.27 pp. At full data the same ten genres give
header-free gains of +1.31, +1.34, and +1.30 pp for IA3, full
fine-tuning, and LoRA (three-seed averages, Section 6.1), so the
matched-size gains, from runs that also used 5 epochs instead of 8 and
funk LoRA at rank 4, lie within 0.1 pp of these, and BitFit's full-data
deficit in the three largest corpora (+0.85 pp over the ten genres)
does not appear at the matched size. Like the full-data gains, these
are about the size of F1's marker handicap (Section 5.3).

The practical reading is that no method is robustly dominant: once the
header targets are set aside, the four methods give the same gain at
matched size, as LoRA, IA3, prefix tuning, and full fine-tuning do
outside Bach chorales at full data, and the all-target rankings reflect
Bach chorale and which runs predicted the C-major key rather than a
data-size or representational effect. This reinforces the main conclusion that the
informative axis is the chord-symbol representation boundary, not a
method leaderboard. (Stage A uses a single seed, so the controlled
comparison is descriptive.)

\begin{figure}
\centering
\includegraphics[width=0.85\linewidth,height=\textheight,keepaspectratio]{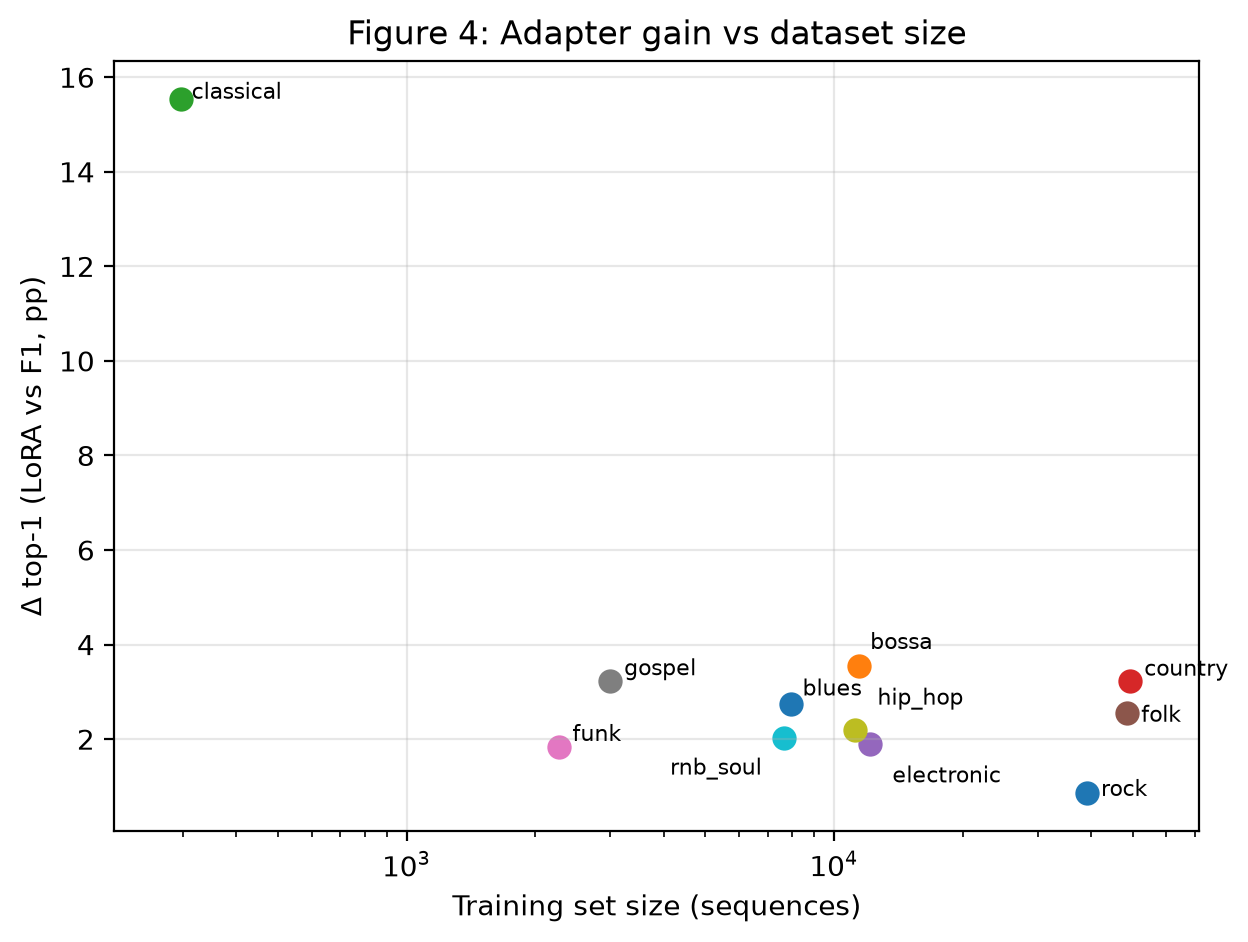}
\caption{Per-genre LoRA gain over F1 versus training-set size, all
targets scored. Outside Bach chorale the gain is about the size of F1's
marker handicap (Section 5.3).}
\end{figure}

\subsubsection{6.11 Epoch-Budget
Sensitivity}\label{epoch-budget-sensitivity}

The main grid fixes an 8-epoch budget. To check that this is sufficient
rather than arbitrary, I sweep 3, 5, 8, and 12 epochs for the three most
data-rich genres (rock, country, folk) under selected-rank LoRA and full
fine-tuning. Best validation loss is essentially flat across the sweep:
rock full fine-tuning holds validation loss at 0.5708 from 3 to 12
epochs, and folk LoRA reaches its minimum (0.5206) by epoch 3.
Meanwhile, training loss keeps falling, indicating mild overfitting
past the early minimum. Because the reported checkpoint is the best-validation one, the
8-epoch budget captures the early optimum and is not a training-length
bottleneck for the method comparison.

\subsubsection{6.12 Decoding Artifacts}\label{decoding-artifacts}

Held-out top-1 measures teacher-forced agreement and does not capture
free-running generation quality. Sampling thirty continuations per genre
from a neutral header prompt and scanning for pathologies gives a mixed
picture. Adaptation removes some failure modes: mean repeat-collapse
drops from 27.3\% (F1) to 0.0\% (LoRA), special-token leakage from 0.9\%
to 0.0\%, and low-diversity from 70.6\% to 43.6\%. Most of these F1
failures come from the prompt's genre marker, which F1 never read in
training.
On rock, the one target marker F1 has read, F1 shows no repeat collapse
and 43.3\% low diversity, close to the rock adapter (0.0\% and 40.0\%),
and with the same sampler F1 prompted with the pop marker gives 0.0\%
repeat collapse and 46.7\% low diversity over 120 samples, against
23.3\% and 80.0\% with the untrained funk marker. But adaptation introduces
another failure: premature end-of-sequence rises from 1.5\% (F1) to 13.3\%
(LoRA), reaching 76.7\% for the Bach-chorale adapter. Improved
teacher-forced accuracy therefore does not eliminate decoding artifacts.
Deployment still needs grammar-aware decoding and post-generation
validation, and top-1 should not be read as musical quality.

\begin{center}\rule{0.5\linewidth}{0.5pt}\end{center}

\subsection{7. Discussion}\label{discussion}

\subsubsection{7.1 What Chord-Symbol Adaptation Can
Do}\label{what-chord-symbol-adaptation-can-do}

The results show that chord-symbol adaptation is useful, within limits.
Across 11 genres and multiple adaptation mechanisms, small interfaces
improve held-out target-genre chord prediction over the frozen base as
it is scored here. Outside Bach chorales, though, the base given the pop
marker matches them without the header targets (Section 5.3), so there
the measurable benefit is a usable genre marker rather than new
harmonic knowledge, while on Bach chorales, whose harmony departs from
the base's training data, LoRA adds 9.6 pp beyond that.
Because the released base is weight-identical to the Phase-0 pop
baseline (Section 4.1), every such gain is a gain over a pop-dominated
harmonic prior. Where the target harmony departs from the base's training data, as
for Bach chorales, these gains mean that genre-local harmonic priors in
the chord-symbol layer can be reached without training a new model from
scratch for every genre.

This is practically important for interactive composition systems. A
user-facing chord tool benefits from modular adapters: genres can be
added, replaced, or disabled without changing the full base model. Even
if a control token is competitive in accuracy, adapters may still be
valuable for deployment, storage isolation, model versioning, and
per-genre update cycles.

\subsubsection{7.2 What It Cannot Do}\label{what-it-cannot-do}

The same results also show the boundary. Method differences are not
decisive. Control-token conditioning is strong. LoRA rank scaling is
shallow in most non-chorale genres. The chord-only classifier is above
chance but weak. Generated LoRA outputs are closer to training
distributions but often lower in entropy. Wrong-genre adapters frequently beat F1 even when they are not matched
to the eval genre, although not F1 given the pop marker.

Together, these findings argue against a strong claim that chord symbols
alone encode genre identity. They encode part of it: harmonic priors,
common transitions, cadential habits, and chord-quality distributions.
They do not encode groove, timbre, voicing, instrumentation, lyrical
idiom, or production. Those missing layers are likely essential for
perceived genre authenticity.

\subsubsection{7.3 Why This Is Not a LoRA
Leaderboard}\label{why-this-is-not-a-lora-leaderboard}

LoRA was the first method tried because it matched the system need:
detachable genre modules over a frozen base. But the expanded experiment
shows that the main scientific object is not LoRA superiority. IA3,
prefix tuning, full fine-tuning, BitFit, and control-token learning all
expose related effects. This is why the paper is framed as a
representation-boundary study. The matched-data-size control makes the
point concrete: when the ten non-chorale genres are sub-sampled to a
common corpus size, the four methods differ by at most about one key
step in each genre (about 0.7 pp in the macro), and that spread is the key
token (Section 6.10). Without the header targets they agree within
0.03 pp there, and outside Bach chorales the methods other than BitFit
agree within 0.05 pp at full data as well, so LoRA's lead in the
11-genre macro comes from Bach chorale and is not evidence of a
superior adapter. In
v2 the base-checkpoint comparison is a same-weights control (Section
6.9), so it no longer speaks to base choice. The right scale for the
small method gaps above is the three-seed spread of each method's
macro, 0.09 to 0.32 pp (Section 6.9).

The most robust sentence supported by the current results is:

\begin{quote}
Chord-symbol adaptation improves genre-local harmonic prediction where
the target harmony departs from the base's training data, as for Bach
chorales, but for genres the base already covers it mainly supplies a
usable genre marker, and chord symbols carry bounded rather than
complete genre identity.
\end{quote}

\begin{center}\rule{0.5\linewidth}{0.5pt}\end{center}

\subsection{8. Limitations and Future
Work}\label{limitations-and-future-work}

First, the evaluation is automatic. Top-1 and top-5 chord prediction
measure agreement with held-out corpus tokens, not musical quality,
usability, or perceived genre authenticity. The real-song subset helps,
but it is not a listening study. This is the main missing validation for
any title or claim involving genre identity.

Second, the corpus is repetitive, and the frozen base has seen most
test songs. Exact duplicates between a target subset's own splits are
rare outside Bach chorales (Section 4.3), but near-duplicate harmonic
patterns are extremely common, and 79.7 to 82.8\% of each non-chorale
test set is in the Phase-0 training split with the same chord sequence
(Section 4.1). This limits generalization claims. A low-overlap or novel-progression
evaluation should become a required robustness check in future work.

Third, genre labels are coarse. A Chordonomicon-derived ``rock'' or
``country'' split covers many substyles and transcription practices.
Bach chorale is not comparable to the pop-adjacent genres and should be
interpreted as a tonal-chorale outlier.

Fourth, chord symbols omit key musical layers. Rhythm, voicing, texture,
production, and timbre can dominate perceived genre identity. Future
work should evaluate whether chord-symbol gains correspond to perceived
genre appropriateness when progressions are rendered or inspected by
musicians.

Fifth, the base-checkpoint ablation in Section 6.9 is, as established
in v2, a same-weights control: the released F1 checkpoint is
weight-identical to the Phase-0 pop baseline, so the comparison
provides a noise calibration but no evidence about base-robustness. A
genuine base ablation (re-running adaptation from an actually
jazz-adapted checkpoint, from an independently pretrained base, or
contrasting adapters with genre-specific from-scratch models)
remains future work, to show whether the representation-boundary result
depends on this particular model family. A suitable jazz-adapted base
now exists: a selection-corrected retrain of the F1 slot (jazz-only
validation selection, released as \texttt{ft-pop80-v2}) is hash-distinct
from Phase-0 (jazz top-1 75.05 on the unified 9-source test). Its
jazz-val selection rule reproduced across three matched-data retrain
seeds, so the blocking factor for this ablation is now experimental
effort rather than checkpoint availability.

Sixth, absolute top-1 depends on tokenization. The flat root-quality
vocabulary collapses enharmonic spellings and all voicing distinctions,
so the reported accuracies are specific to this chord-symbol scheme. A
finer or coarser vocabulary would shift them. LoRA rank, target modules,
and other adapter hyperparameters were only lightly swept and may also
move the method-level numbers.

Seventh, the deployed composition tool re-ranks adapter outputs with
additional physics- and theory-retrieval modules (R1/R2) and renders
multi-track arrangements with rule-based voicing (R3). This report
isolates the chord-symbol language-model contribution. Those rerank and
arrangement layers, and voice-leading beyond the root-position symbols
the model emits, are post-processing outside the present scope and are
left to future work.

Two additional comparisons would make the same claim more defensible.
A genre-conditioned n-gram, Markov, or retrieval chord model would
establish a lower bound for local transition statistics, and a direct
comparison with style-token chord generation systems such as
Chordinator would clarify how this work differs from them. The code,
adapters, and evaluation scripts are now public (Section 5.4), so the
representation-boundary claim can be checked and both comparisons can
be run on the same pipeline.

\begin{center}\rule{0.5\linewidth}{0.5pt}\end{center}

\subsection{9. Conclusion}\label{conclusion}

This report studies chord-symbol time-series adaptation as an
interpretable middle layer for controllable music AI. The results are
positive but bounded. Multiple adaptation methods improve a frozen
chord model across eleven target genres as it is scored here, but
outside Bach chorales the frozen model given the pop marker does as well
without the header tokens. Matched adapters beat wrong-genre ones on
average, but outside Bach chorales their edge over the best wrong
adapter is within about 0.2 pp once the header tokens are excluded. As documented in v2, the released
base weights are those of the Phase-0 pop baseline, so every gain
is a gain over a pop-dominated harmonic prior. However, no method is decisively
superior after correction (and once the header tokens are set aside, the methods give the same
gain outside Bach chorales, apart from BitFit on the three largest
corpora, where a test-time reload most likely degraded its cells), while
control-token conditioning is strong, chord-only genre classification
remains weak, and generated-output diagnostics show distribution
matching rather than open-ended diversity.

The best interpretation is therefore not that LoRA solves genre
adaptation, nor that chord symbols capture full genre identity. The
evidence supports a narrower and more useful conclusion: chord-symbol
sequences carry measurable harmonic genre information, enough to support
modular adaptation, but not enough to replace rhythm, timbre,
arrangement, and human perceptual evaluation. The most reusable findings are that small conditioning over a shared
harmonic base matters more than the exact adapter family, and that a
frozen base has to be scored with a genre marker it was trained on,
since an unseen marker alone costs it 1.0 to 1.9 pp in every
non-chorale genre but rock. This makes chord symbols a
useful controllable layer, and also marks the boundary that the next
stage of the research must cross.

\protect\phantomsection\label{refs}

\end{document}